*Electric-Field-Induced Phase Separation and Labyrinthine Patterns in Nanocolloids.*

Xiaodong Duan and Weili Luo

Department of Physics, Advanced Materials Processing and Analysis Center, Center for Drug Discovery and Diagnostics, University of Central Florida, Orlando, FL 32816

Brent Wacaser and Robert C. Davis

Dept. of Physics and Astronomy
Brigham Young University
Provo, Utah, 84604

Abstract

**In this paper we report universal labyrinthine patterns observed in nanocolloids originated from electric-field-induced phase separation. For nanocolloids consisting of magnetic particles (magnetic fluids), the labyrinthine pattern as the result of the co-existence of two phases is found to be stable against perturbation by an additional magnetic field. The relaxation back to the globally disordered pattern after the magnetic field is removed clearly shows evolution of "undulation" thus demonstrates that superimposing magnetic and electric fields can easily manipulate the patterns resulted from the phase separation, indicating that a magnetic fluid has an advantage in exploring the dynamics of defects that has applications in many fields.**

Recently, in the field of pattern formation, increased activities have been carried out in colloidal systems because colloids, besides being model systems for studying some fundamental issues in physics such as two-dimensional melting[7], glassy behaviors[8], and field-induced instabilities[9], have diverse applications in oil, food, cosmetic, chemical, automobile (Electro- and Magneto-Rheological fluids have found applications in shock absorbers and clutch fluids), and biotech industries[10]. By common definition that pattern formation refers to the process in which ordered structures occurs in an initially *homogeneous* system, patterns formed in colloids can be divided into two types—one originates from energetic competitions in which entropy does not play an important role. Examples include Eletro- and Magneto-Rheological (ER or M R) fluids consisting of micron-sized particles[11, 12]. The other type involves hard sphere systems where the internal energy does not change with the particle configurations thus only entropy dominates[13]. In this paper, we discuss a new mechanism for external-field-driven pattern formation in nanocolloids originated from the competition between the internal energy and the entropy. The resulting structure is a labyrinthine pattern that has been seen in many diverse systems but *with different mechanism*. Because of small size of particles in nanocolloids (10nm instead of microns), the Brownian motion or the entropic contribution to the free energy is as important as that of internal energy, a mechanism rarely studied before. We will demonstrate that a nanocolloid or a colloidal system consisting of *nanoparticles* is a model system for studying thermodynamics of interacting systems. By comparing two very different systems, one consists magnetic particles the other pure dielectric nano spheres, the observed labyrithine patterns seems to be universal. Nevertheless, a magnetic colloid or magnetic fluid (MF) have special advantage since the morphology of patterns can be controlled by an additional magnetic field. The mechanism discussed here should be general for any colloid, including biological, physical, and chemical systems, where the internal energy and entropy terms in the free energy are comparable in values. We want to emphasize that the well-known labyrinthine pattern in MF first observed by Rosensweig, Zahn, and R. Schumovich sometime ago[6] was done in a MF surrounded by *another fluid* that is immiscible with MF and the labyrinthine pattern appeared when a *magnetic* field is applied. Mechanism responsible for that phenomenon[14] is very different from the one to be discussed here.

The first nanocolloid studied here is a magnetic fluid consisting of magnetite ($Fe_3O_4$) nanoparticles suspended in kerosene [15] with each particle coated by a surfactant layer of 2nm thick. The mean diameter of particles is 10nm, the average magnetic moment for each particle 2 x $10^4$ $\mu_B$, with $\mu_B$ the Borh magneton, and the volume concentration of the magnetite particles 7%. Since the flipping rate of the magnetic moment in a particle along its easy axis is about 1 GHz at room temperature, the time average of the magnetic moment is almost zero without external magnetic field. A thin layer of magnetic fluid with thickness of 10$\mu$m is sandwiched between two glass plates with area of 1$cm^2$ whose inner sides are coated with ITO (Indium-Tin Oxide) with conductivity of $10^6$ S/m. The conductivity of the magnetic fluid is in the order of $10^{-7}$S/m, much smaller than that of ITO conducting layer, so we can treat the conducting layer in an electric field that is perpendicular to the sample cell as a surface with an equal potential. The relative dielectric constants of kerosene and $Fe_3O_4$ particles are 2.7 and 15, respectively [16]. The patterns in the magnetic fluid could be observed by an optical microscope and recorded by digital CCD camera.

In zero and small applied voltage, $V_a$, the system is in its original homogenous phase, as shown in fig.1 (a). When $V_a$ is larger than a critical value, $V_{ac}$ = 7.5V, phase separation occurs (fig.1 (b)-(d)): the two phases that are denser (smaller transmission to light) and more dilute (higher transmission to light) than the original phase form labyrinthine pattern in a time scale of 10 seconds. When increasing the applied voltage the light contrast between the two phases increases while the pattern remains static, suggesting that the concentration difference between the two phases increases as well. However, the average width of the stripes and their spacing almost remain constant within our time scale. The process is reversible when the field is switched off. We can find the concentration distribution in the system, in principle, by measuring the light transmission in each point in the image. The intensity of the transmitted light could be expressed as: $I = I_0 \exp(-\alpha L)$, where $I_0$ is the intensity of the incident light, $L$ the thickness of magnetic fluid, and $\alpha$ the absorption coefficient. The absorption of light in magnetic fluid, $\alpha$, mainly comes from that of particles, $\alpha \approx \alpha_0 f$, with $f$ the volume fraction of the particles and $\alpha_0$ $\approx 0.83 \mu m^{-1}$. For our thin sample, $I$ could be approximately expanded as $I \approx I_0(1-\alpha_0 L f)$. If we apply this expression in both the concentrated and diluted regions in the phase separated sample, and use $f = (f_+ + f_-)/2$ with $f_+$ and $f_-$ the concentration for the dense and the dilute phases, respectively, we get the average transmitted light intensity as: $I = (I_+ + I_-)/2 \approx I_0(1-\alpha_0 L f)$, $I_+$ and $I_-$ represent the transmitted light intensity in concentrated and diluted regions, respectively. Here we assume the areas for the dense and dilute regions are the same, based on the experimental observation. Therefore, to the first order approximation, the total transmitted light intensity will be unchanged with phase separation if the incident light intensity is kept a constant, which is verified in our experiment as shown in the inset of fig.2. It indicates that $I \approx I_0(1-\alpha_0 L f)$ is a good approximation. We define the contrast, $c$, as $c = (I_+ - I_-)/(I_+ + I_-)$ thus $c \propto (f_+ - f_-)$. By using the BioScan-Optima software, we can digitize light intensity in locations in our images shown in fig.1. We plot contrast c as a function of square of the electric field, i.e., c vs. $E^2$, in fig.2. The threshold can be clearly identified.

Although the contrast seems to approach the saturation, The well-established labyrinthine pattern is still evolving with time. We have measured

Since the dielectric constant for a particle, $\varepsilon_p$, is higher than that of carrier fluid, $\varepsilon_f$, each particle obtains an induced dipole moment in an external electric field: $p = ba^3 \varepsilon_f E_l$, where mismatch factor $b = (\varepsilon_p - \varepsilon_f)/(\varepsilon_p + 2\varepsilon_f)$, a is the radius of the particle, $E_l$ the local electric field [17]. We introduce a dimensionless quantity $\lambda = k_B T d^3 / p^2$ that characterizes the competition between the thermal energy and the dipolar interaction. Here d = 2a is the diameter of the particle. Using the typical parameters for our magnetic fluid, $\lambda \cong 10-100$ in the electric field range studied, suggesting that thermal energy must be considered when discussing phase separation in our system. The free energy for the induced dipole system could be expressed as: $F = U_E - TS$, where $U_E$ is the total electrostatic energy due to dipole interactions induced by electric field. For a dielectrics confined between two conductors in constant electric field, $U_E$ is $\int dV \{\varepsilon_o [-\varepsilon_e + \varepsilon_p f + \varepsilon_f (1-f)] E^2 / 2\}$, where $\varepsilon_e$ is the effective dielectric constant of the magnetic fluid. For dilute binary systems ($\phi < 0.2$), Lichteneker has proposed a phenomenological description [18]:

$$\varepsilon_e = \exp[(1-f)\ln\varepsilon_f + f\ln\varepsilon_p] \qquad (1)$$

which works well for disordered media.

The entropy term in the free energy, -TS, favors the homogenous phase because the larger the entropy the lower the free energy. If the competition between energy and entropy could be observed experimentally, they should have the same order of magnitude in value change resulted from the phase separation. When $f \ll 1$ and $\lambda > 1$, the entropy per unit volume could be expressed as [19, 20]:

$$S = -k_B \frac{f}{v}(\ln f - 1) \qquad (2)$$

where $v$ is the correlated volume. If the particles are completely independent, $v$ is just the particle volume [19, 20]. However, if particles have strong interaction or aggregate into clusters, they can not be treated as independent particles. In such a situation, we use the correlated volume instead of the particle volume to calculate the entropy.

In the phase separated state, if the volume fraction of the dense phase is $\gamma$, the dilute phase is 1-$\gamma$, then the expression of the free energy per unit volume, $f$, will be:

$$f = -1/2 \{(g[e(f_+) - f_+ e_P - (1-f_+)e_f] + (1-g)[e(f_-) - f_- e_P - (1-f_-)e_f])e_o E^2\}$$
$$- k_B T \{g[\frac{f_+}{n}(\ln f_+ - 1)] + (1-g)[\frac{f_-}{n}(\ln f_- - 1)]\}$$

$$(3)$$

where $\phi_+$, $\phi_-$ represent the volume fractions of dense and diluted phases, respectively. And the total volume fraction of particles is $f = gf_+ + (1-g)f_-$. In our experiment, the regions of the dense and dilute phases have almost the same areas, so we simply take $\gamma = 0.5$. The condition for minimum free energy requires:

$$\frac{1}{2}\{e_o e_e(f) E^2 (\ln \frac{e_P}{e_f})(f_+ - f_-)\} = \frac{k_B T}{n} \ln \frac{f_+}{f_-} \qquad (4)$$

We define the ratio $r$ as

$$r = \frac{e_o e_e (\ln \frac{e_P}{e_f})^2 E^2}{2k_B T / n} \qquad (5)$$

then we get the equation of phase separation.

$$r = \frac{\ln f_+ - \ln f_-}{f_+ - f_-} \quad \text{and} \quad f_+ + f_- = 2f \qquad (6)$$

This equation has nontrivial solution only if $r\phi > 1$. For our system, $\phi = 0.07$, so r must be greater than the critical value $r_c = 14.3$, indicating the existence of a critical field, $E_c$, above which phase separation occurs. In fig.3 we show the solution of the phase separation equation for our magnetic fluid, plotted as the particle volume fractions in dense (solid line) and the dilute (dashed line) regions as a function of control parameter r. It shows that the total volume fraction is indeed a constant. Now we can calculate the contrast between the two phases, c, as a function of r or the applied field, E. In fig.2 the solid line is the result of the theoretical calculation. We found that the saturation electric field $E_s$ is about 1.4 times the critical field $E_c$, agreeing with experiment very well where $E_s = 11V$ and $E_c = 7.5V$. The extracted $v$ is larger than the volume for the single particle, indicating particle correlation or aggregation in the applied electric field. Except for ideal gas, the theory for entropy of interacting systems is a difficult task. The agreement between theory and experiment shown in fig.2 suggests that the entropy expression in eq. (2) may be applicable in an interacting system with a suitable $v$.

The resulting pattern of the phase separation is labyrinthine. This type of electric-field-induced labyrinths has not been studied in colloids before. The phenomenon is different from the labyrinthine patterns in magnetic fluids discussed in literature where *magnetic fluid and another fluid immiscible with it* form labyrinths upon applying a *magnetic field due to a totally different mechanism*[6, 14]. While in other colloids such as ER fluids, where electric field does induce phase separation, in most cases the resulting patterns are columnar rather than labyrinthine because in those systems entropy is not important [11, 12]. Although deviation from columnar structure was observed in ER fluids, the fibrillated morphology is quite different from the labyrinth shown in this work [21]. In MR fluids or similar systems, highly ramified structure have been observed and some authors have draw similarities between them with labyrinthine structures. These branched structures, however, is completely different type where micron sized particles form solid, branched networks while here the labyrinthine morphology is formed between two liquid phases that have different concentrations. The mechanism for "branches" in MR fluids[21] or concentrated magnetic fluids that phase separates in *zero field* [2], originates from dipolar repulsions between fictitious surface charges at the end of columns where systems encounter substrates. The alternating "stripe" phase induced by electric field in nanocolloids is "thermodynamic", i.e. it arises from entropy of the system. From above equations, we realize that the entropy term strongly depends on particle size. If the particle size is 1μm instead of nanometers as in nanocolloids, the $v$ will increase and the entropy will decrease by several orders of magnitude, then the electrostatic energy will dominate the phase separation, as in electrorheological fluids. If the particle size is much smaller, for example, in the order of several angstroms, the free energy change due to the entropy will be increased by four orders of magnitude, thus dominates the electrostatic energy. As the result there will be no phase separation in reasonable field range. The other extreme is a hard sphere system in which the internal energy does not depend on the configurations of particle arrangement so the free energy for configurational change is solely determined by entropy, again a different mechanism from phenomenon discussed here.

This new mechanism should be general, independent particle types. In order to test this hypothesis and verify our conclusion that the labyrinthine pattern is from competition between electrostatic energy and entropy rather than the magnetic moments of particles, we conducted similar experiment on another nanocolloid consisting of 10 nm CdSe particles coated with surfactants dispersed in hexane. Similar phase separation and labyrinthine patterns were also found as shown in fig.4 with the same sample thickness of 16 μm and the same applied voltages. Therefore, the observed phase separation and the resulting pattern is independent of particle types as long as they have similar dimensions. Due to magnetic moments of particles, however, the labyrinthine patterns in magnetic fluids can be controlled by an additional magnetic field, B. In fig. 5, we demonstrate the effect of superimposing a magnetic field, B, which is in the sample plane and is perpendicular to the applied electric field. When a magnetic field of 25 Gauss is applied, the stripes are stretched along the B field direction with some dislocations remain. After switching off the magnetic field, the straight parallel stripes relax back to the globally disordered labyrinthine pattern gradually, indicating that the labyrinth represents a stable state. The relaxation process is also interesting. Initially, the long parallel stripes are disturbed by the thermal motion and start to undulate. Then the structure becomes more disordered by continued deforming around the topological defects as shown in fig. 5. Some of the stripes break open and new topological defects emerge. Following Seul and Wolfe [3], we introduce a parameter, $\kappa = d_{ee}/l$, defined as the ratio of end to end distance $d_{ee}$ to contour length l. $\kappa \sim 1$ represents the lamellae and $\kappa \ll 1$ indicates the highly random labyrinthine configuration. In fig.6 we show the evolution of deformation of the lamellae to the stable disordered state. As a consequence, in magnetic fluid, manipulating patterns formed in electric field is made easier by additional magnetic field, making magnetic fluid unique and more interesting.

We have demonstrated that electric-field-induced phase separation and the resulting labyrinthine pattern is *universal in nanocolloids*. The underlying physics is attributed to the competition between the internal energy and entropy. The entropy behaves as "repulsion" that causes stripes alternating in particle concentration in the resulting structure. We further showed that superimposing a magnetic field on top of electric field provides a useful way to study dynamics of topological defects that has application in diverse fields such as controlling domain orientation in block copolymer thin film [22], nanolithography [23], langmuir monolayers [24], and magnetic garnet films for bubble technology [25].

Figure Captions

Fig.1 Phase separation process in a magnetic fluid induced by DC electric field. The applied voltages are (a) 7.10 V; (b) 7.64 V; (c) 8.35 V; (d) 9.82 V.

Fig.2 Transmission contrast, c, versus electric field. ■ experimental data, which is proportional to the volume fraction difference between concentrated and dilute regions. The solid line is the theoretical result for volume fraction contrast: $(\varphi_+ - \varphi_-)/(\varphi_+ + \varphi_-)$ multiplied by a constant to convert it to transmission contrast to compare with experimental $c$. The inset is the average transmission light density in different applied voltages.

Fig.3 The solution of equation (6) for phase separation in electric field. The solid line and dash line represent the particle volume fraction of concentrated region $\varphi_+$ and that of dilute region $\varphi_-$, respectively. The control parameter $r$ is defined in the text.

Fig.4 Labyrinthine pattern resulted from phase separation in applied voltage of 16 V on the 16 μm thickness samples of (a) a magnetic fluid and (b) a colloid consisting of 10nm CdSe particles dispersed in hexane.

Fig.5 (a) Labyrinthine pattern in applied voltage of 10 V. (b) when a magnetic field of 25 Gauss is applied, "bended" labyrinth is stretched to a nearly lamella configuration with remaining defects. After the magnetic field is switched off the relaxation back to original globally disordered labyrinth is observed as a function of time: (c) t = 22s; (d) t = 52s.

Fig.6 Time dependence of aspect ratio for the stripe pattern $\langle d_{ee}/l \rangle$, where $d_{ee}$ is the end-to-end distance and $l$ the contour length, following the definition in ref.[3]. $\langle d_{ee}/l \rangle \sim 1$ corresponds to lamellae and $\langle d_{ee}/l \rangle \ll 1$ corresponds to disordered labyrinthine patterns.

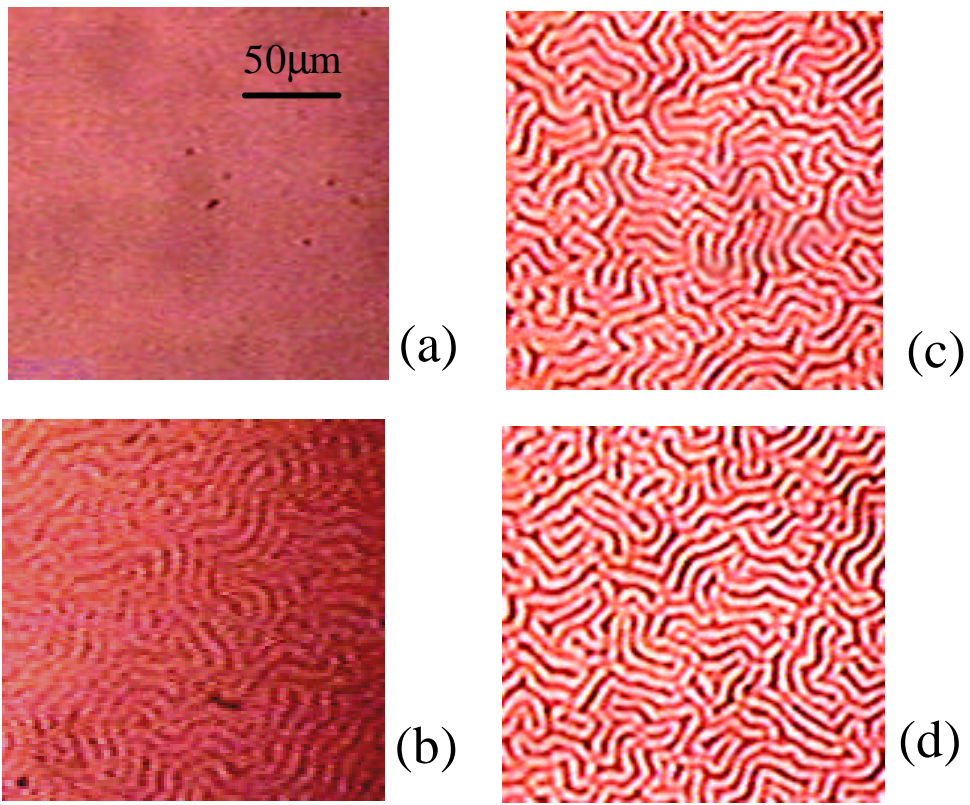

Xiaodong Duan, Weili Luo, Brent Wacaser, and Robert C. Davis *Electric-field-induced phase separation*… Figure 1

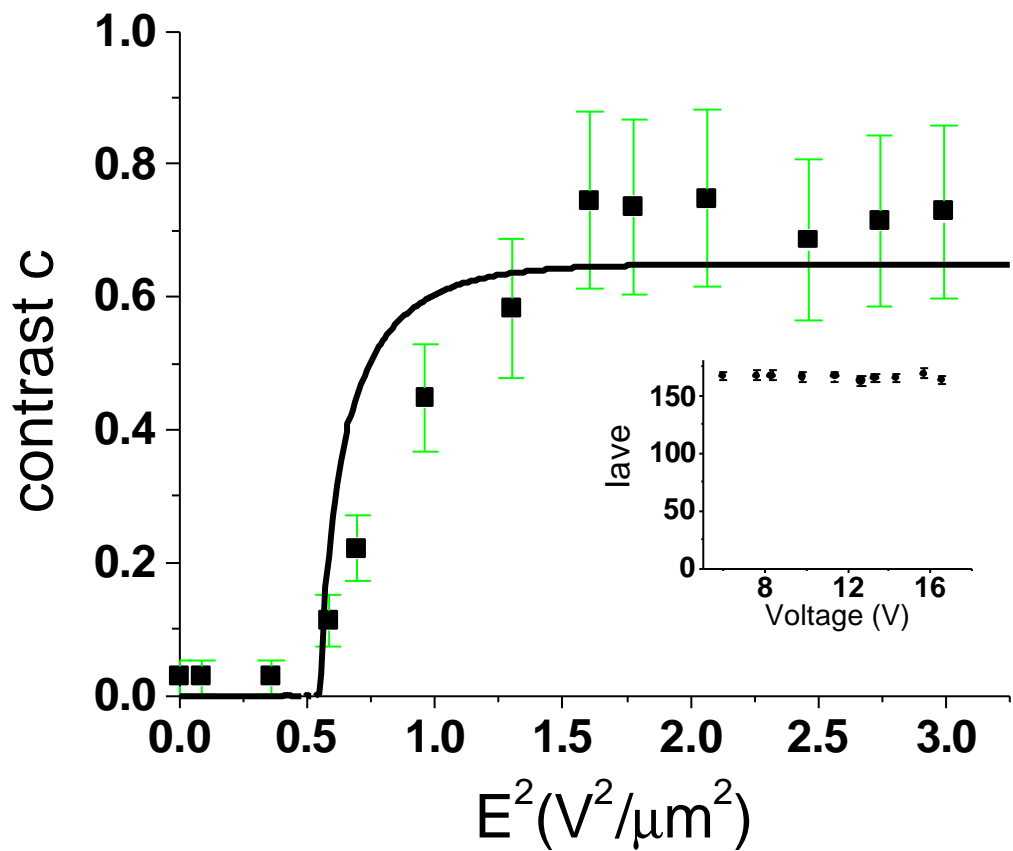

(a) 50mm  (b) 50mm

Xiaodong Duan, Weili Luo, Brent Wacaser, and Robert C. Davis, *Electric-field-induced phase separation…* Figure 2

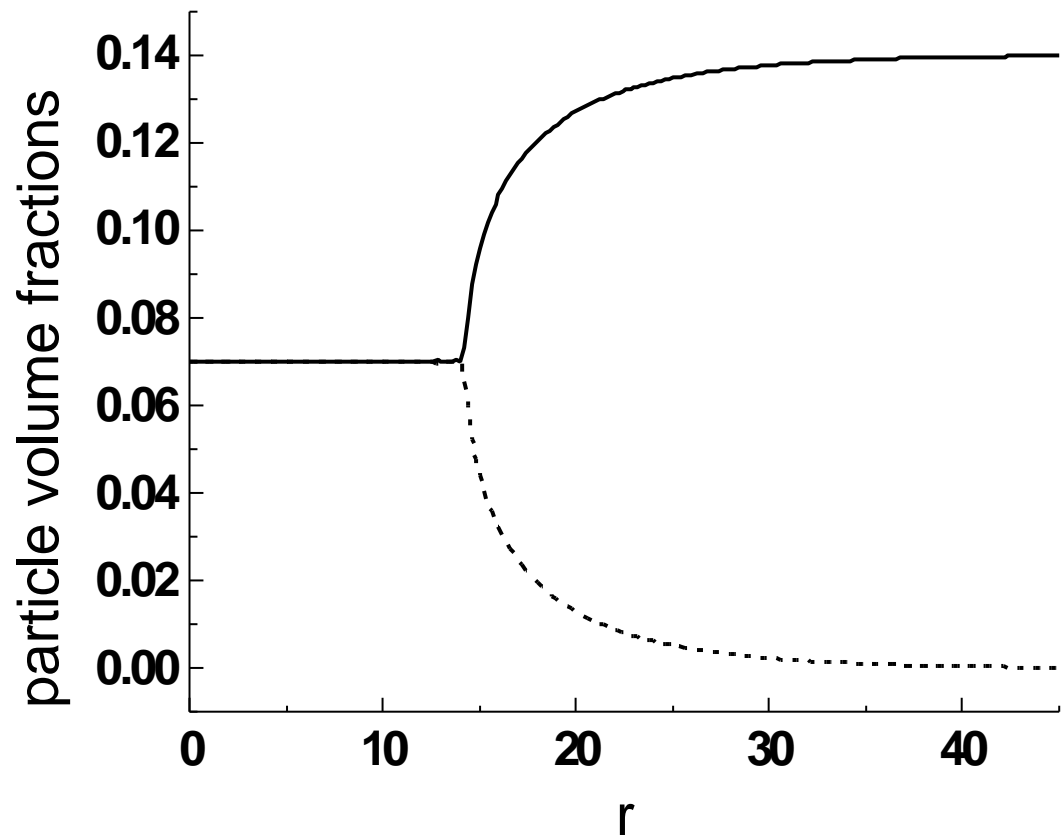

Duan, Luo, Wacaser, and Davis, *Electric-Field-Induced Universal Labyrinthine ...* Figure 3

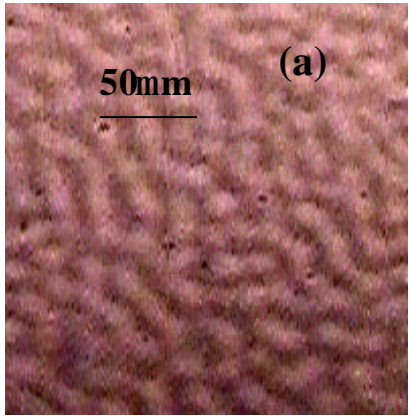 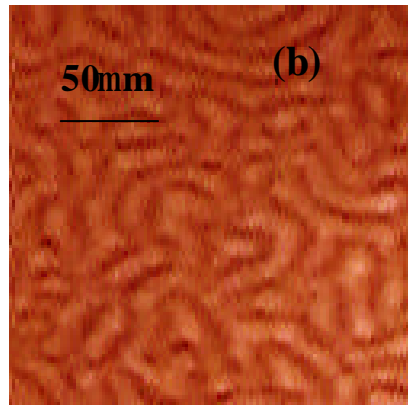

Duan, Luo, Wacaser, and Davis, *Electric-Field-Induced Universal Labyrinthine...* Figure 4

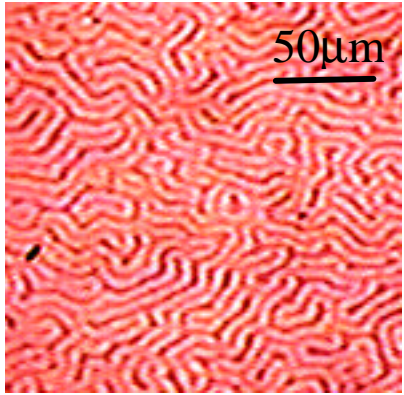
(a)

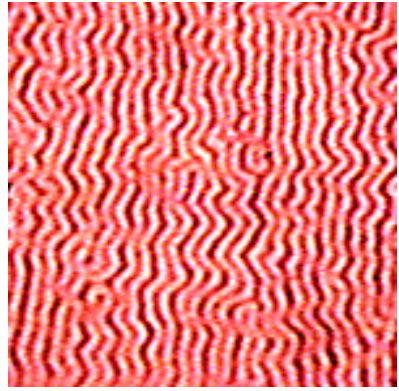
(c)

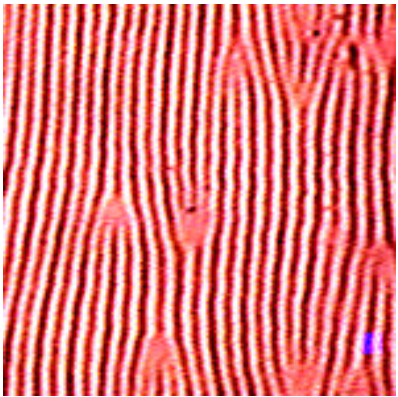
(b) ↑B

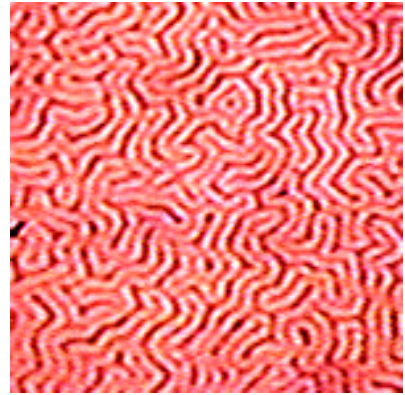
(d)

Duan, Luo, Wacaser, and Davis, *Electric-field-induced phase separation*… Figure 5

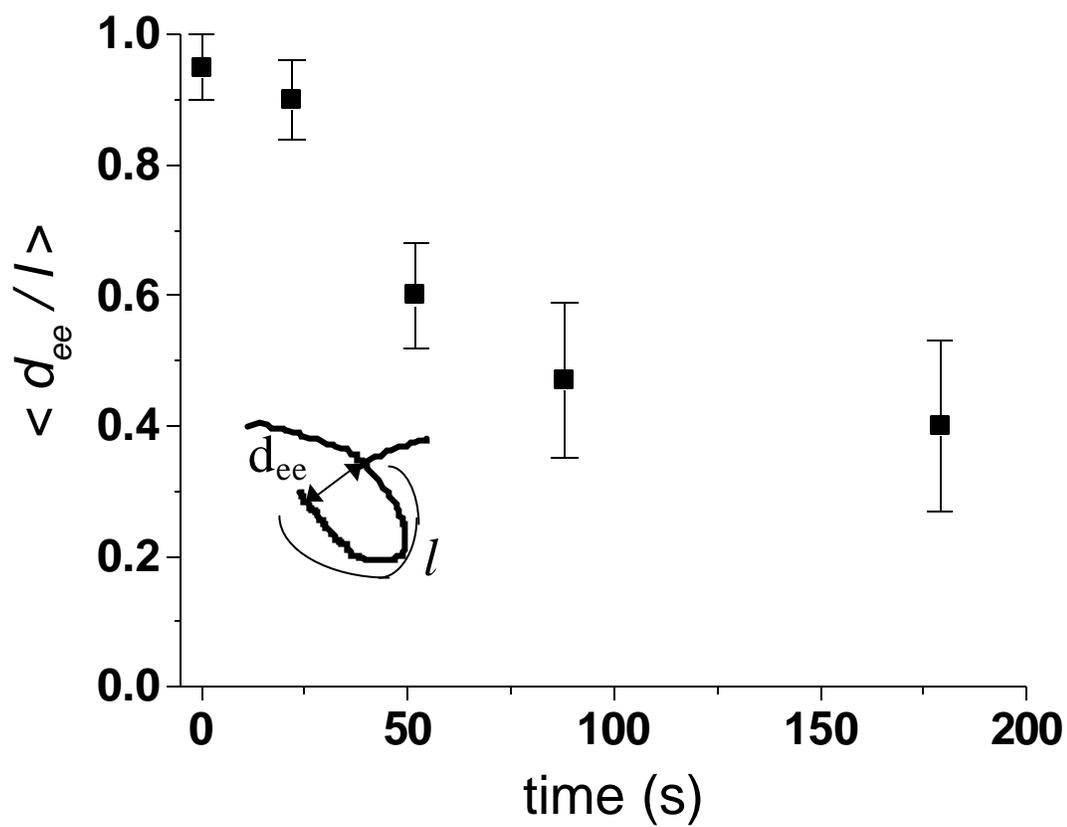

Duan, Luo, Wacaser, and Davis, *Electric-field-induced phase separation*… Figure 6